\documentclass[final,authoryear,5p,times,twocolumn]{elsarticle}

\usepackage{graphicx}
\usepackage[usenames]{color} 
\usepackage{amssymb}
\usepackage{hyperref}
\usepackage[english]{babel}

\journal{Journal of Theoretical Biology}

\newcommand{\av}[1]{\langle {#1} \rangle}

\begin{document}

\begin{frontmatter}
  
\title{Immunization strategies for epidemic processes in
time-varying contact networks
}

  \author[label1]{Michele Starnini}
  \author[label2,label3]{Anna Machens}
  \author[label4]{Ciro Cattuto}
  \author[label2,label3,label4]{Alain Barrat}
  \author[label1]{Romualdo Pastor-Satorras}
  
  \address[label1]{Departament de F\'\i sica i Enginyeria Nuclear, Universitat
    Polit\`ecnica de Catalunya, Campus Nord B4, 08034 Barcelona,
    Spain}
  \address[label2]{Aix Marseille Universit\'e, CNRS, CPT, UMR 7332, 13288 Marseille, France}
  \address[label3]{Universit\'e de Toulon, CNRS, CPT, UMR 7332, 83957 La Garde, France}
  \address[label4]{Data Science Laboratory, ISI Foundation, Torino, Italy}

\begin{abstract}
  Spreading processes represent a very efficient tool to investigate
  the structural properties of networks and the relative importance of
  their constituents, and have been widely used to this aim in static
  networks. Here we consider simple disease spreading processes on
  empirical time-varying networks of contacts between individuals, and
  compare the effect of several immunization strategies on these
  processes. An immunization strategy is defined as the choice of a
  set of nodes (individuals) who cannot catch nor transmit the
  disease.  This choice is performed according to a certain ranking of
  the nodes of the contact network. We consider various ranking
  strategies, focusing in particular on the role of the training
  window during which the nodes' properties are measured in the
  time-varying network: longer training windows correspond to a larger
  amount of information collected and could be expected to result in
  better performances of the immunization strategies. We find instead
  an unexpected saturation in the efficiency of strategies based on
  nodes' characteristics when the length of the training window is
  increased, showing that a limited amount of information on the
  contact patterns is sufficient to design efficient immunization
  strategies. This finding is balanced by the large variations of the
  contact patterns, which strongly alter the importance of nodes from
  one period to the next and therefore significantly limit the
  efficiency of any strategy based on an importance ranking of
  nodes. We also observe that the efficiency of strategies that
  include an element of randomness and are based on temporally local
  information do not perform as well but are largely independent on
  the amount of information available.

\end{abstract}
  
  
\end{frontmatter}

\section{Introduction}
\label{sec:introduction}

The topology of the pattern of contacts between individuals plays a
fundamental role in determining the spreading patterns of epidemic
processes \citep{keeling05:_networ}. The first predictions of
classical epidemiology \citep{anderson92,cite-keeling2007mid} were
based on the homogeneous mixing hypothesis, assuming that all
individuals have the same chance to interact with other individuals in
the population. This assumption and the corresponding results were
challenged by the empirical discovery that the contacts within
populations are better described in terms of networks with a
non-trivial structure \citep{Newman2010}. Subsequent studies were
devoted to understanding the impact of network structure on the
properties of the spreading process. The main result obtained
concerned the large susceptibility to epidemic spread shown by
networks with a strongly heterogeneous connectivity pattern, as
measured by a heavy-tailed degree distribution $P(k)$ (defined as the
probability distribution of observing one individual connected to $k$
others) with a a diverging second
moment~\citep{pv01a,Lloyd18052001,newman02,refId0}.

The original studies considered the interaction networks as static
entities, in which connections are frozen or evolve at a time scale
much longer than the one of the epidemic process.  This static view of
interaction networks hides however the fact that connections appear,
disappear, or are rewired on various timescales, corresponding to the
creation and termination of relations between pairs of
individuals~\citep{Holme:2011fk}. Longitudinal data has traditionally
been scarce in social network analysis, but, thanks to recent
technological advances, researchers are now in a position to gather
data describing the contacts in groups of individuals at several
temporal and spatial scales and resolutions.

The analysis of empirical data on several types of human interactions
(corresponding in particular to phone communications or physical
proximity) has unveiled the presence of complex temporal patterns in
these systems
\citep{Hui:2005,PhysRevE.71.046119,Onnela:2007,10.1371/journal.pone.0011596,Tang:2010,Stehle:2011nx,Miritello:2011,Karsai:2011,Holme:2011fk}.
In particular, the heterogeneity and burstiness of the contact
patterns are revealed by the study of the distribution of the
durations of contacts between pairs of agents, the distribution of the
total time in contact of pairs of agents, and the distribution of gap
times between two consecutive interactions involving a common
individual.  All these distributions
are indeed heavy-tailed (often compatible with power-law behaviors),
which corresponds to the burstiness of human
interactions~\citep{barabasi2005origin}.

These findings have led to a large modeling effort
\citep{Scherrer:2008,Hill:2009,Gautreau:2009,PhysRevE.81.035101,PhysRevE.83.056109,2012arXiv1203.5351P,starnini_modeling_2013}
and stimulated the study of the impact of a network's dynamics on the
dynamical processes taking place on top of it. The processes studied
in this context include synchronization~\citep{albert2011sync},
percolation~\citep{Parshani:2010,Bajardi:2012}, social
consensus~\citep{consensus_temporal_nrets_2012}, or
diffusion~\citep{PhysRevE.85.056115}. Epidemic-like processes have
also been explored, both using realistic and toy models of propagation
processes
\citep{Rocha:2010,Isella:2011,Stehle:2011nx,Karsai:2011,Miritello:2011,dynnetkaski2011,Panisson:2012,Holme:2013,Rocha:2013,Masuda13}.
The study of simple schematic spreading processes over temporal
networks helps indeed expose several properties of their dynamical
structure: dynamical processes can in this context be conceived as
probing tools of the network's temporal structure \citep{Karsai:2011}.

The study of spreading patterns on networks is naturally complemented
by the formulation of vaccination strategies tailored to the
specific topological (and temporal) properties of each network.
Optimal strategies shed light on how the role and importance of nodes
depend on their properties, and can yield importance rankings of
nodes. In the case of static networks, this issue has been
particularly stimulated by the fact that heterogeneous networks with a
heavy-tailed degree distribution have a very large susceptibility to
epidemic processes, as represented by a vanishingly small epidemic
threshold. In such networks, the simplest strategy consisting in
randomly immunizing a fraction of the nodes is ineffective. More
complex strategies, in which nodes with the largest number of
connections are immunized, turn out to be effective
\citep{PhysRevE.65.036104} but rely on the global knowledge of the
network's topology. This issue is solved by the so-called acquaintance
immunization~\citep{PhysRevLett.91.247901}, which prescribes the
immunization of randomly chosen neighbors of randomly chosen
individuals.

Few works have addressed the issue of the design of immunization
strategies and their respective efficiency in the case of dynamical
networks~\citep{Lee:2010fk,Tang:2011,Takaguchi:2012,Masuda13}.  In
particular, \cite{Lee:2010fk} consider datasets describing the
contacts occurring in a population during a time interval $[0,T]$;
they define and study strategies that use information from the
interval $[0,\Delta T]$ to decide which individuals should be
immunized in order to limit the spread during the remaining time
$[\Delta T, T]$.  Specifically, the authors introduce two strategies,
called \textit{Weight} and \textit{Recent}. In the Weight strategy, a
fraction $f$ of nodes is selected at random: for each of these nodes,
his/her most frequent contact in the interval $[0,\Delta T]$ is
immunized. In the Recent strategy, the last contact before $\Delta T$
of each of the randomly chosen individuals is immunized.  Both
strategies are defined in the spirit of the acquaintance immunization,
insofar as they select nodes using only partial (local) information on
the network. Using a large $\Delta T=75\% T$, \cite{Lee:2010fk} show
that these strategies perform better than random immunization and show
that this is related to the temporal correlations of the dynamical
networks.

In this paper, we investigate several immunization strategies in
temporal networks, including the ones considered by \cite{Lee:2010fk},
and address in particular the issue of the length $\Delta T$ of the
``training window'', which is highly relevant in the context of
real-time, specific tailored strategies. The scenario we have in mind
is indeed the possibility to implement a real-time immunization
strategy for an ongoing social event, in which the set of individuals
to the immunized is determined by strategies based on preliminary
measurements up to a given time $\Delta T$. The immunization problem
takes thus a two-fold perspective: The specific rules (strategy) to
implement, and the interval of time over which preliminary data are
collected. Obviously, a very large $\Delta T$ will lead to more
complete information, and a more satisfactory performance for most
targeting strategies, but it incurs in the cost of a lengthy data
collection. On the other hand, a
short $\Delta T$ will be cost effective, but yield a smaller amount of
information about the observed social dynamics.

In order to investigate the role of the training window length on the
efficiency of several immunization strategies, we consider a simple
snowball susceptible-infected (SI) model of epidemic spreading or
information diffusion~\citep{anderson92}. In this model, individuals
can be either in the susceptible (S) state, indicating that they have
not been reached by the ``infection'' (or information), or they can be
in the infectious (I) state, meaning that they have been infected by
the disease (or that they have received the information) and can
further propagate it to other individuals.  Infected individuals do
not recover, i.e., once they transition to the infectious state they
remain indefinitely in that state.  Despite its simplicity, this model
has indeed proven to provide interesting insights into the temporal
structure and properties of temporal networks. Here we focus on the
dynamics of the SI model over empirical time-varying social
networks. The networks we consider describe time-resolved face-to-face
contacts of individuals in different environments and were measured by
the SocioPatterns collaboration
(\texttt{http://www.sociopatterns.org}) using wearable proximity
sensors~\citep{10.1371/journal.pone.0011596}.  We consider the effect
on the spread of an SI model of the immunization of a fraction of
nodes, chosen according to different strategies based on different
amounts of information on the contact sequence.  We find a saturation
effect in the increase of the efficiency of strategies based on nodes
characteristics when the length of the training window is
increased. The efficiency of strategies that include an element of
randomness and are based on temporally local information do not
perform as well but are largely independent on the amount of
information available.

The paper is organized as follows: we briefly describe the empirical
data in Sec. \ref{sec:empir-cont-sequ}. In
Sec. \ref{sec:epid-models-numer} we define the spreading model and
some quantities of interest. The immunization strategies we consider
are listed in Sec. \ref{sec:immun-strat}.
Sec. \ref{sec:numerical-results} contains the main numerical results,
and we discuss in Sec. \ref{sec:effects-temp-corr} and
Sec. \ref{sec:effects-non-determ} the respective effects of temporal
correlations and of randomness effects in the spreading model. Section
\ref{sec:conclusions} finally concludes with a discussion on our
results.

\section{Empirical contact sequences}
\label{sec:empir-cont-sequ}

We consider temporal networks describing the face-to-face close
proximity of individuals in different contexts, collected by the
SocioPatterns collaboration. We refer to
\texttt{http://www.sociopatterns.org} and
\citep{10.1371/journal.pone.0011596} for details on the data collection
strategy, which is based on wearable sensors worn by individuals.  The
datasets give access, for each pair of participating individuals, to
the list of time intervals in which they were in face-to-face close
proximity ($\approx 1-2$m), with a temporal resolution of $20$
seconds.

In this paper, we use temporal social networks measured in several
different social contexts: the 2010 European Semantic Web Conference
(``eswc''), a geriatric ward of a hospital in Lyon (``hosp''), the
2009 ACM Hypertext conference (``ht''), and the 2009 congress of the
Soci\'et\'e Francaise d'Hygi\`ene Hospitali\`ere (``sfhh'').  These
data correspond therefore to the fast dynamics of human contacts over
the scale of a few days.  A more detailed description of the
corresponding contexts and analyses of these datasets can be found in
\citep{10.1371/journal.pone.0011596,percol,Isella:2011,Stehle:2011,Panisson:2012}.
In Table~\ref{tab:summary} we summarize some of properties of the
considered datasets.

\begin{table}[b]
  \begin{center}
    \begin{tabular}{||c||c|c|c|c|c||}
      \hline
      Dataset & $N$ & $T$ & $ \av{k}$ &  $ \av{s}$ & $\bar{f}$ \\ \hline 
      eswc &  173  & 4703   & 50  & 370  &  6.8\\
      ht      &  113 & 5093   & 39  & 366  &  4.1 \\
      hosp &   84  & 20338 & 30  & 1145 &  2.4   \\
      sfhh  &  416 & 3834   & 54  & 502  & 27.2\\ 
      \hline
    \end{tabular}
  \end{center}
  \caption{Some properties of the SocioPatterns datasets under
    consideration: number of different individuals engaged in
    interactions ($N$); 
    total duration of the contact sequence ($T$), measured in
    intervals of length $\Delta t =20$ sec.;
    average degree $\av{k}$ (number of different contacts) and average
    strength $\av{s}$ (total time spent in face-to-face interactions)
    of the network of contacts aggregated over the whole sequence; 
    average number of interactions $\bar{f}$ at each time step.}
  \label{tab:summary}
\end{table}

\section{Epidemic models and numerical methods}
\label{sec:epid-models-numer}

We simulate numerically the susceptible-infected (SI) spreading
dynamics on the above describe datasets of human face-to-face
proximity.  The process is initiated by a single infected individual
(``seed'').  At each time step, each infected individual $i$ infects
with probability $\beta$ the susceptible individuals $j$ with whom $i$
is in contact during that time step.  The process stops either when
all nodes are infected or at the end of the temporal sequence of
contacts.

Different individuals have different contact patterns and \textit{a
  priori} contribute differently to the spreading process. In order to
quantify the spreading efficiency of a given node $i$, we proceed as
follows: We consider $i$ as the seed of the SI process, all other
nodes being susceptible. We measure the half prevalence time, i.e.,
the time $t_i$ needed to reach a fraction of infected nodes equal to
$50 \%$ of the population.  Since not all nodes appear simultaneously
at $t=0$ of the contact sequence, we define the \textit{half-infection
  time} of seed node $i$ as $T_i = t_i - t_{0,i}$, where $t_{0,i}$ is
the time at which node $i$ first appears in the contact sequence. The
half-infection time $T_i$ can thus be seen as a measure of the spreading power of
node $i$: smaller $T_i$ values correspond to more efficient spreading
patterns.

\begin{figure}[t]
  \begin{center}
    \includegraphics[clip=true,trim=0cm 0cm 0cm 0cm,width=8.5cm]{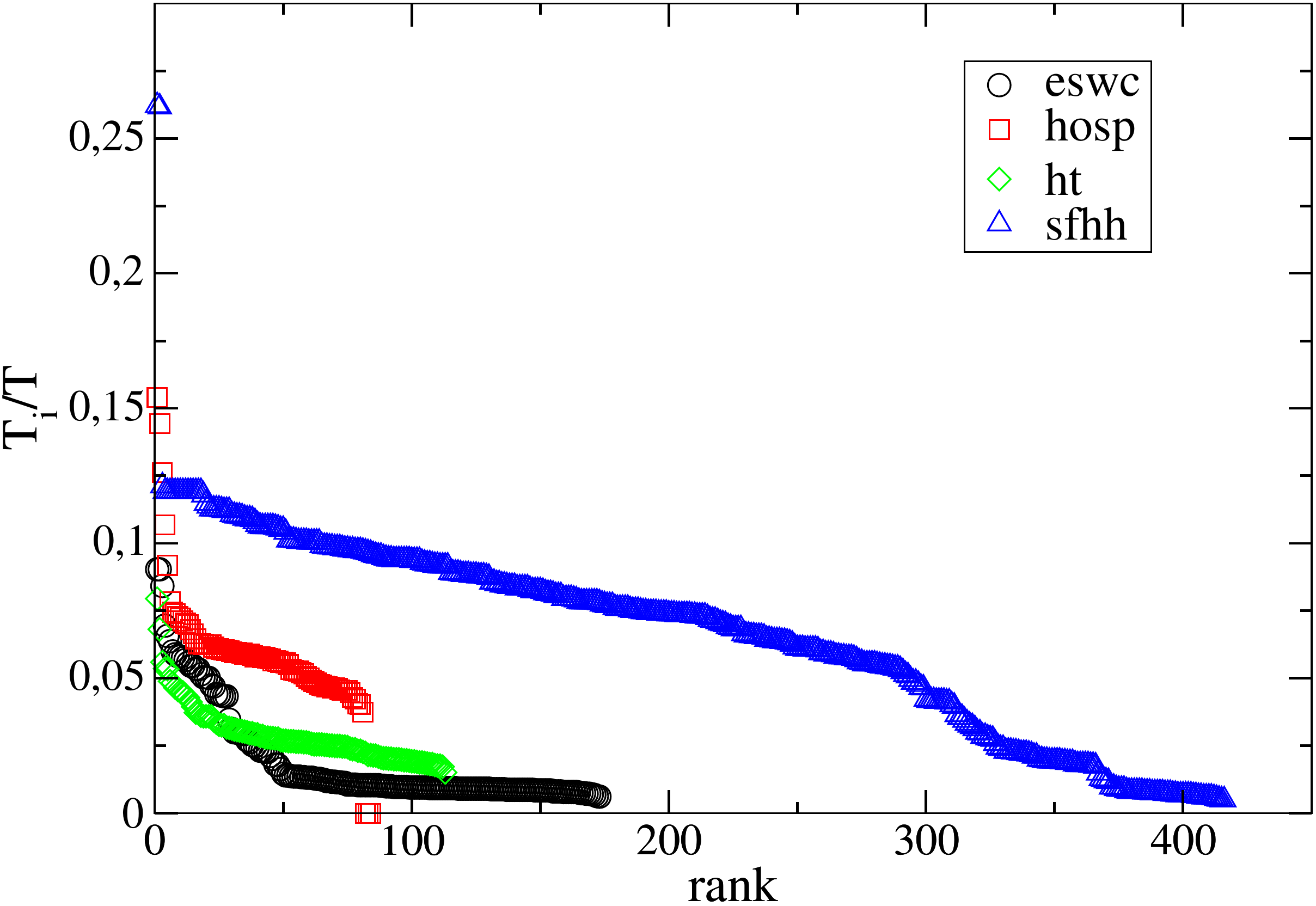}
  \end{center}
  \caption{Rank plot of the half-infection times $T_i$ divided by the
    contact sequence duration $T$ for the various datasets.}
  \label{fig:rank_t}
\end{figure}

We first focus on the deterministic case $\beta=1$ (the effects of
stochasticity, as given by $\beta<1$, are explored in
Sec.~\ref{sec:effects-non-determ}).  Figure~\ref{fig:rank_t} shows
rank plot of the rescaled half-infection times $T_i/T$ for various
datasets, where $T$ is the duration of the contact sequence.  We note
that $T_i$ is quite heterogeneous, ranging from $T_i < 5\% T$ up to
$T_i \simeq 20 \% T$ \footnote{We note that defining $T_i$ as the time
  needed to reach a different fraction of the population, such as
  e.g. $25\%$, leads to a similar heterogeneity.}.

Some nodes are therefore much more efficient spreaders than
others. This implies that the immunization of different nodes could
have very different impacts on the spreading process. To estimate this
impact, we define for each node $i$ the infection delay ratio $\tau_i$
as
\begin{equation}
\label{eq:1}
\tau_i = \left\langle \frac{ T_j^{i}- T_j }{T_j} \right\rangle_{j \neq i} ,
\end{equation}
where $T_j^{i} $ is the half-infection time obtained when node $j$ is
the seed of the spreading process and node $i$ is immunized, and the
ratio is averaged over all possible seeds $j \neq i$ and over
different starting times for the SI process (the half-infection time
being much smaller than the total duration of the contact sequence,
$T_j \simeq 0.1 T$) \footnote{Note that in some cases, node $i$ is not
  present during the time window in which the SI process is simulated;
  in this case, $T_j^{i}= T_j$.}.  The infection delay ratio $\tau_i$
quantifies therefore the average impact that the immunization of node
$i$ has on SI processes unfolding over the temporal network.

\begin{figure}[t]
\begin{center}
\includegraphics*[clip=true,trim=0cm 1cm 0cm 1cm,width=8.5cm]{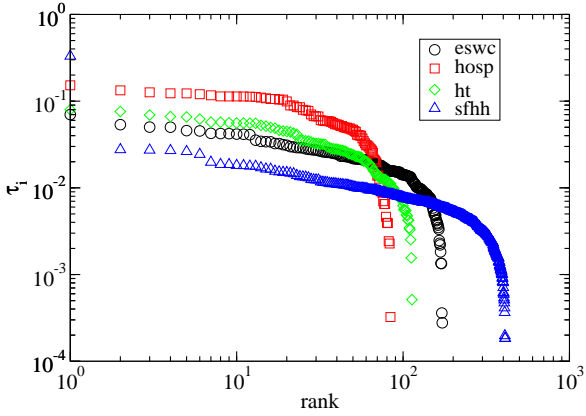} 
\end{center}
\caption{Rank plot of the infection delay ratio $\tau_i$ for various
  datasets.}
  \label{fig:rank_tau}
\end{figure}

Figure \ref{fig:rank_tau} displays a rank plot of $\tau_i$ for various
datasets. As expected, the immunization of a single node does most
often lead to a limited delay of the spreading dynamics.
Interestingly however, $\tau_i$ is broadly distributed and large
values are also observed.

The infection delay ratio of a single node $i$, $\tau_i$, can be
generalized to the case of the immunization of any set of nodes
$\mathcal{V} =\{i_1,\ldots, i_n \}$, with $n<N$.  We measure the
spreading slowing down obtained when immunizing the set $\mathcal{V}$
through the infection delay ratio
\begin{equation}
\label{eq:2}
\tau_{\mathcal{V}} = \left\langle \frac{ T_j ^\mathcal{V} - T_j }{T_j}
\right\rangle_{j \notin \mathcal{V}} , 
\end{equation}
where $T_j^\mathcal{V}$ is the half-infection times of node $j$ when
all the nodes of set $\mathcal{V}$ are immunized, and the average is
performed over all possible seeds $j \notin \mathcal{V}$ and different
starting times for the SI process.

In addition to slowing down the propagation process, the immunization
of certain individuals can also block the spreading paths towards
other, non-immunized, individuals, limiting in this way the final
number of infected individuals. We measure this effect through the
\textit{average outbreak size ratio}
\begin{equation}
\label{eq:3}
i_\mathcal{V} = - \left\langle \frac{ I_j ^\mathcal{V} - I_j }{I_j}
\right\rangle_{j \notin \mathcal{V}} , 
\end{equation}
where $I^\mathcal{V}_j$ and $I_j$ are the number of infected
individuals (outbreak size) for an SI process with seed $j$, with and
without immunization of the set $\mathcal{V}$, respectively.  The
ratio is averaged over all possible seeds $j \notin \mathcal{V}$ and
over different starting times of the SI process.

\section{Immunization strategies}
\label{sec:immun-strat}

An immunization strategy is defined by the choice of the set
$\mathcal{V}$ of nodes to be immunized.  We define here different
strategies, and we compare their efficiencies in section
\ref{sec:numerical-results} by measuring $\tau_\mathcal{V}$ and
$i_\mathcal{V}$.  More precisely, for each contact sequence of
duration $T$ we consider an initial temporal window $[0, \Delta T]$
over which various properties of nodes can be measured.  A fraction
$f$ of the nodes, chosen according to different possible rules, is
then selected and immunized (it forms the set $\mathcal{V}$). 
Finally, $\tau_\mathcal{V}$ and
$i_\mathcal{V}$ are computed by simulating the SI process with and
without immunization and averaging over starting seeds and times.  For
each selection rule, the two relevant parameters are $f$ and $\Delta
T$.  Larger fractions $f$ are naturally expected to lead to larger
$\tau_\mathcal{V}$ and $i_\mathcal{V}$.  Here we also consider the
effect of $\Delta T$, where a larger $\Delta T$ corresponds to a
larger amount of information on the contact sequence.  We investigate
whether and how more information about the contact sequence yields a
higher efficiency of the immunization strategy.

We consider the following strategies (or "protocols"):
\begin{itemize}
\item[\textbf{K}] Degree protocol. We immunize the $f N$ individuals
  with the highest aggregated degree in $[0, \Delta T]$
  \citep{PhysRevE.65.036104}; the aggregated degree of an individual
  $i$ corresponds to the number of different other individuals with
  whom $i$ has been in contact during $[0, \Delta T]$;

\item[\textbf{BC}] Betwennness centrality protocol. We immunize the $f
  N$ individuals with the highest betweenness centrality measured on
  the aggregated network in $[0, \Delta T]$
  \citep{PhysRevE.65.056109};
  
\item[\textbf{A}] Acquaintance protocol. We choose randomly an
  individual and immunize one of his contacts in $[0,\Delta T]$,
  repeating the process until $f N$ individuals are immunized
  \citep{PhysRevLett.91.247901};

\item[\textbf{W}] Weight protocol. We choose randomly an individual and
  immunize his most frequent contact in $[0,\Delta T]$, repeating for
  various elements until $f N$ individuals are immunized
  \citep{Lee:2010fk};

\item[\textbf{R}] Recent protocol. We choose randomly an individual and
  immunize his last contact in $[0,\Delta T]$, repeating for various
  elements until $f N$ individuals are immunized \citep{Lee:2010fk}.

\end{itemize}

As a benchmark, we also consider the following two strategies:
\begin{itemize}
\item[\textbf{Rn} ] Random protocol. We immunize $f N$ individuals
  chosen randomly among all nodes;

\item[\textbf{T}] $\tau$-protocol. We immunize the first $f N$
  individuals with the highest $\tau_i$, with $\tau_i$ calculated
  according to Eq.~(\ref{eq:1}) in the interval $[0,\Delta T]$.
\end{itemize}
The \textbf{Rn} strategy uses no information about the contact
sequence and we use it as a worst case performance baseline.  The
\textbf{T} strategy makes use, through the quantity $\tau_i$, of the
entire information about the contact sequence as well as complete
information about the average effect of node immunization on SI
processes taking place over the contact sequence. It could thus be
expected to yield the best performance among all strategies.

The \textbf{A}, \textbf{W}, \textbf{R} and \textbf{Rn} strategies
involve a random choice of individuals.  In each of these cases, we
average the results over $10^2$ independent runs (each run
corresponding to an independent choice of the individuals to
immunize).

\section{Numerical results}
\label{sec:numerical-results}

\begin{figure}[t]
\begin{center}
\includegraphics*[clip=true,trim=0cm 0cm 0cm 0.6cm,width=8.5cm]{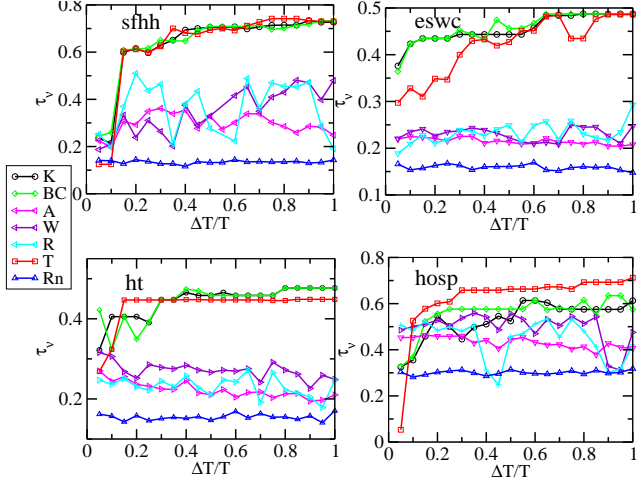}
\end{center}
\caption{Infection delay ratio $\tau_{\mathcal{V}}$ as a function of
  the training window $\Delta T$ for different immunization protocols,
  for various datasets.  The fraction of immunized individuals is
  $f=0.05$.}
  \label{fig:delay_dt}
\end{figure}

\begin{figure}[t]
\begin{center}
\includegraphics*[clip=true,trim=0cm 0cm 0cm 0.6cm,width=8.5cm]{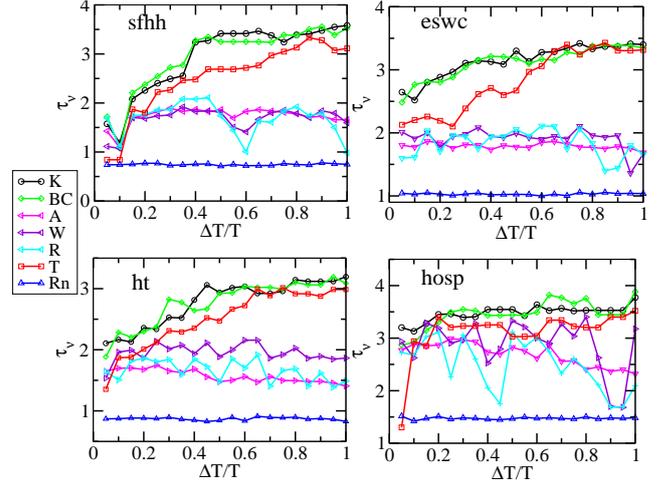}
\end{center}
\caption{Infection delay ratio $\tau_{\mathcal{V}}$ as a function of
  the training window $\Delta T$ for different immunization protocols,
  for various datasets.  The fraction of immunized individuals is
  $f=0.2$.}
  \label{fig:delay_dt_f02}
\end{figure}

We first study the role of the temporal window $\Delta T$ on the
efficiency of the various immunization strategies. To this aim, we
consider two values of the fraction of immunized individuals, $f=0.05$
and $f=0.2$, and compute the infection delay ratio $\tau$ as a
function of $\Delta T$ for each immunization protocol, and for each
dataset.  The results, displayed in Figs.~\ref{fig:delay_dt} and
\ref{fig:delay_dt_f02}, show that an increase in the amount of
information available, as measured by an increase in $\Delta T$, does
not necessarily translate into an larger efficiency of the
immunization, as quantified by the delay of the epidemic process.  The
$\textbf{A}$, $\textbf{W}$ and $\textbf{R}$ protocols have in all
cases lower efficiencies that remain almost independent on $\Delta
T$. Moreover, and in contrast with the results of \cite{Lee:2010fk} on
a different dataset, $\textbf{W}$ and $\textbf{R}$ do not perform
better than $\textbf{A}$.  On the other hand, the immunization
efficiency of the $\textbf{K}$, $\textbf{BC}$ and $\textbf{T}$
protocols increases at small $\Delta T$ and reaches larger values for
all the datasets. As expected, the \textbf{Rn} protocol, which does not use any
information, fares the worst. For $f=0.05$, all protocols yield an
infection delay ratio that is largely independent from $\Delta T$ for
large enough training windows $\Delta T \gtrsim 0.2 T$.  For $f=0.2$,
the increase of $\tau$ is more gradual but tends to saturate for
$\Delta T \gtrsim 0.4 T$ as well.  In all cases, a limited knowledge
of the contact time series is therefore sufficient to estimate which
nodes have to be immunized in order to delay the spreading dynamics,
especially for small $f$, i.e., in case of limited resources.
Interestingly, in some cases, the $\textbf{K}$ and $\textbf{BC}$
protocols lead to a larger delay of the spread than the $\textbf{T}$
protocol, despite the fact that the latter is designed to explicitly
identify the nodes which yield the maximal (individual) infection
delay ratio.  This could be ascribed to correlations between the
activity patterns of nodes, leading to a non-linear dependence on $f$
of the immunization efficiency (in particular, the list of nodes to
immunize is built using the list of degrees, betweenness centralities,
and $\tau_i$ values computed on the original network, without
recomputing the rankings each time a node is removed).

\begin{figure}[t]
\begin{center}
\includegraphics*[clip=true,trim=0cm 0.8cm 0cm 0.8cm,width=8.5cm]{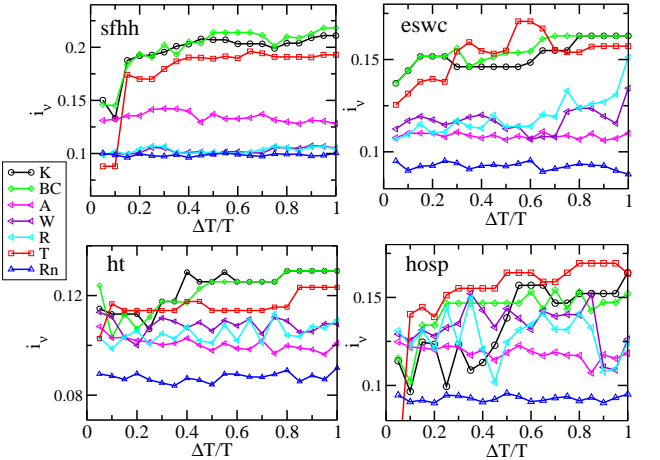}
\end{center}
\caption{Average outbreak ratio $i_{\mathcal{V}}$ as a function of the temporal
  window $\Delta T$ for different vaccination protocols, for various
  datasets.  Here the fraction of immunized nodes is $f=0.05$.}
  \label{fig:infect_dt}
\end{figure} 

Figure \ref{fig:infect_dt} reports the outbreak ratio
$i_{\mathcal{V}}$ as a function of the temporal window $\Delta T$ for
different vaccination protocols. Results similar to the case of the
infection delay ratio are recovered: the reduction in outbreak size,
as quantified by the average outbreak size ratio defined in
Eq. (\ref{eq:3}), reaches larger values for the degree, betweenness
centrality and $\textbf{T}$ protocols than for the $\textbf{A}$,
$\textbf{W}$ and $\textbf{R}$ protocols.

\begin{figure}[t]
\begin{center}
\includegraphics*[clip=true,trim=0cm 0.5cm 0cm 0.8cm,width=8.5cm]{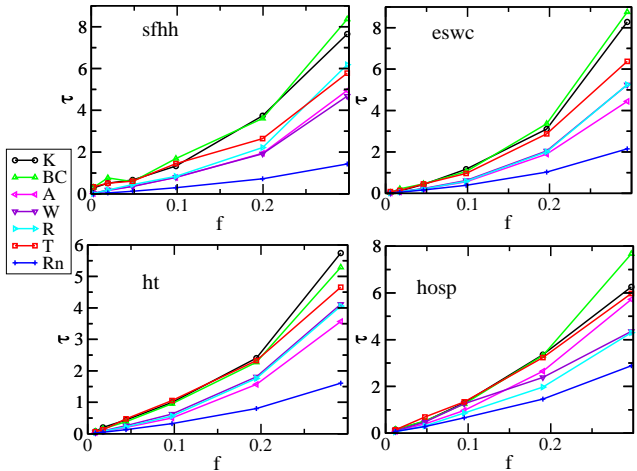}
\end{center}
\caption{Infection delay ratio $\tau_{\mathcal{V}}$ as a function of
  the fraction $f$ of immunized elements, for different vaccination
  protocols, for various datasets, and a fixed $\Delta T=0.4 T$. }
  \label{fig:delay_f}
\end{figure}

\begin{figure}[t]
\begin{center}
\includegraphics*[clip=true,trim=0cm 0.5cm 0cm 0.8cm,width=8.5cm]{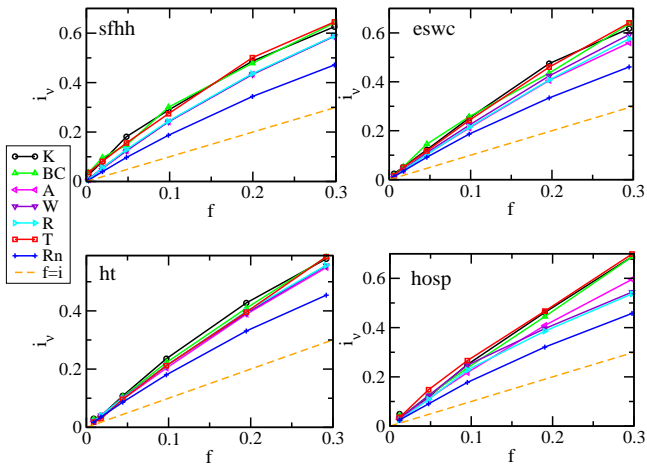}
\end{center}
\caption{Average outbreak ratio $i_{\mathcal{V}}$ as a function of the
  fraction $f$ of immunized elements, for different vaccination
  protocols, for various datasets. The dashed line represents the
  fraction of immunized individuals, $f$. Here $\Delta T=0.4 T$. }
  \label{fig:infect_f}
\end{figure}

We finally investigate the robustness of our results when the fraction
of immunized individuals varies. To this aim, we use a fixed length
$\Delta T=0.4 T$ for the training window and we plot the infection
delay ratio $\tau_{\mathcal{V}}$ and the average outbreak size ratio
$i_{\mathcal{V}}$ as a function of $f$, respectively, in
Figs. \ref{fig:delay_f} and \ref{fig:infect_f}.  The results show that
the ranking of the strategies given by these two quantities is indeed
robust with respect to variations in the fraction of immunized
individuals.  In particular, the $\textbf{K}$ and $\textbf{BC}$
protocols perform much better than the $\textbf{W}$ and $\textbf{R}$
protocols for at least one of the efficiency indicators.
 
\section{Effects of temporal correlations}
\label{sec:effects-temp-corr}

Real time-varying networks are characterized by the presence of bursty
behavior and temporal correlations, which impact the unfolding of
dynamical processes
\citep{PhysRevE.71.046119,Kostakos:2009,Isella:2011,Nicosia:2011,Bajardi:2012,Holme:2011fk,PhysRevE.85.056115}.
For instance, if a contact between vertices $i$ and $j$ takes place
only at the (discrete) times ${\cal T}_{ij} \equiv
\{t_{ij}^{(1)},t_{ij}^{(2)},\cdots,t_{ij}^{(n)} \}$, it cannot be used
in the course of a dynamical processes at any time $t \not\in {\cal
  T}_{ij}$.  A propagation process initiated at a given seed might
therefore not be able to reach all the other nodes, but only those
belonging to the seed's set of influence \citep{PhysRevE.71.046119},
i.e., those that can be reached from the seed by a time respecting
path.

In order to investigate the role of temporal correlations, we consider
a reshuffled version of the data in which correlations between
consecutive interaction among individuals are removed. To this aim, we
consider the list of events $(i,j,t)$ describing a contact between $i$
and $j$ at time $t$ and reshuffle at random their time stamps to build
a synthetic uncorrelated new contact sequence. We then apply the same
immunization protocols to this uncorrelated temporal network.

\begin{figure}[t]
  \begin{center}
    \includegraphics[clip=true,trim=0cm 0cm 0cm 0.8cm,width=8.5cm]{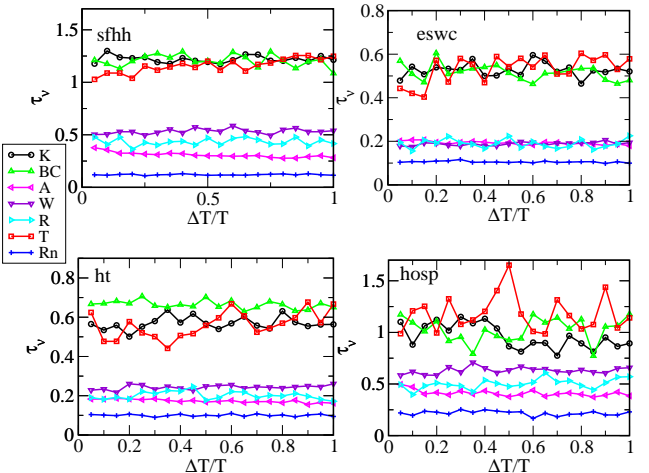}
  \end{center}
  \caption{Infection delay ratio $\tau_{\mathcal{V}}$ as a function of
    the training window length $\Delta T$, computed on one instance of
    a randomized dataset in which the time stamps of the contacts have
    been reshuffled, for different immunization protocols and various
    datasets, with $f=0.05$.  }
  \label{fig:delay_sran}
\end{figure}

Figure \ref{fig:delay_sran} displays the corresponding results for the
infection delay ratio $\tau_{\mathcal{V}}$ computed for SI spreading
simulations performed on a randomized dataset (similar results are
obtained for the average outbreak size ratio $i_{\mathcal{V}}$).  We
have checked that our results hold across different realizations of
the randomization procedure.  The efficiency of the protocol is then
largely independent of the training window length.  As the contact sequence is
random and uncorrelated, all temporal windows are statistically
equivalent, and no new information is added by increasing $\Delta T$:
in particular, as nodes appear in the randomly reshuffled sequence
with a constant probability that depends on their empirical activity,
the ranking of nodes according for instance to their aggregated degree
remains very stable as $\Delta T$ changes, so that a very small
$\Delta T$ is enough to reach a stable such ranking.  Nevertheless,
the efficiency ranking of the different protocols is unchanged: the
degree, betweenness centrality, and $\textbf{T}$ protocols outperform
the other immunization strategies.  Moreover, the efficiency levels
reached are higher than for the original contact sequence: the
correlations present in the data limit the immunization efficiency in
the case of the present datasets.  Note that studies of the role of
temporal correlations on the speeding or slowing down of spreading
processes have led to contrasting results, as discussed by
\cite{Masuda13}, possibly because of the different models and dataset
properties considered.

\section{Non-deterministic spreading}
\label{sec:effects-non-determ}

\begin{figure}[t]
  \begin{center}
    \includegraphics[clip=true,trim=0cm 0cm 0cm 0.8cm,width=8.5cm]{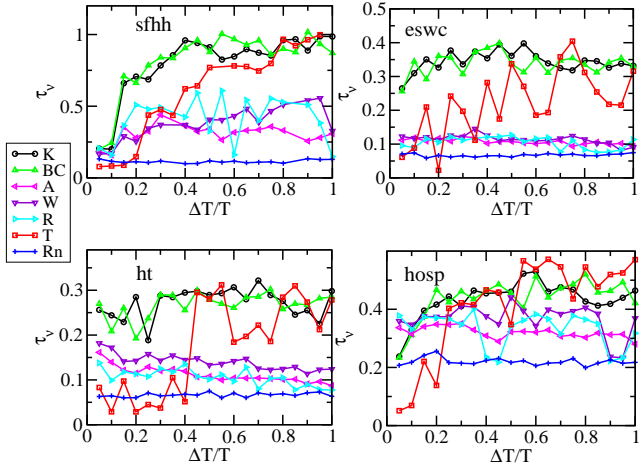}
  \end{center}
  \caption{Infection delay ratio $\tau_{\mathcal{V}}$ as a function of the temporal
    window $\Delta T$ in a probabilistic SI with $\beta=0.2$, for
    different vaccination protocols, various datasets, and $f=0.05$. }
  \label{fig:delay02}
\end{figure}

We also verify the robustness of our results using a probabilistic SI
process with $\beta=0.2$. We consider the same immunization strategies
and we compute the same quantities as in the case $\beta=1$.  Given
the probabilistic nature of the spreading process, we now average the
above observables over $10^2$ realizations of the SI process.  Figure
\ref{fig:delay02} shows that our results hold in the case of a
probabilistic epidemics spreading, although in this case the infection
delay ratio $\tau_{\mathcal{V}}$ presents a noisy behavior, due to the
stochastic fluctuations originated in the probabilistic spreading
dynamics.  The average outbreak ratio $i_{\mathcal{V}}$, not shown,
behaves in a very similar way.  Thus, also in this more realistic case
with $\beta < 1$, a limited knowledge of the contact sequence is
enough to identify which individuals to immunize.

\section{Discussion}
\label{sec:conclusions}

Within the growing body of work concerning temporal networks, few
studies have yet considered the issue of immunization strategies and
of their efficiency. In general terms, the amount of information that
can be extracted from the data at hand about the characteristics of
the nodes and links is a crucial ingredient for the design of optimal
immunization strategies. Understanding how much information is needed
in order to design good (and even optimal) strategies, and how the
efficiency of the strategies depend on the information used, remain
largely an open questions whose answer might depend on the precise
dataset under investigation.

We have here leveraged several datasets describing contact patterns
between individuals in varied contexts, and performed simulations in
order to measure the effect of different immunization strategies on
simple SI spreading processes. We have considered immunization
strategies designed according to different principles, different ways
of using information about the data, and different levels of
randomness.  Strategies range from the completely random $\textbf{Rn}$
to the $\textbf{A}$, $\textbf{W}$ and $\textbf{R}$ strategies that
include a random choice, to the fully deterministic $\textbf{K}$,
$\textbf{BC}$ and $\textbf{T}$ that are based on various node's
characteristics. Moreover, $\textbf{K}$ uses only local information
while $\textbf{BC}$ and $\textbf{T}$ rely on the global knowledge of
the connection patterns.

The strategies that are most efficient, as measured by the change in
the velocity of the spread and by the final number of nodes infected,
are the deterministic protocols, namely $\textbf{K}$ and
$\textbf{BC}$.  Strategies based on random choices, even when they are
designed in order to try to immunize "important" nodes, are less
efficient.

We have moreover investigated how the performance of the various
strategies depends on the time window on which the nodes'
characteristics are measured. A longer time window corresponds indeed
a priori to an increase in the available information and hence to the
possibility to better optimize the strategies. We have found, however,
a clear saturation effect in the efficiency increase of the various
strategies as the training window on which they are designed
increases. This is particularly the case when the fraction of
immunized nodes is small (Fig. \ref{fig:delay_dt}), for which a small
$\Delta T$ is enough to reach saturation, while the saturation is more
gradual for larger fractions of immunized
(Fig. \ref{fig:delay_dt_f02}).  Moreover, the strategies that involve
a random component yield results that are largely independent on the
amount of information considered.

\begin{figure}[t]
  \begin{center}
    \includegraphics[clip=true,trim=0cm 0.5cm 0cm 2.5cm,width=9cm]{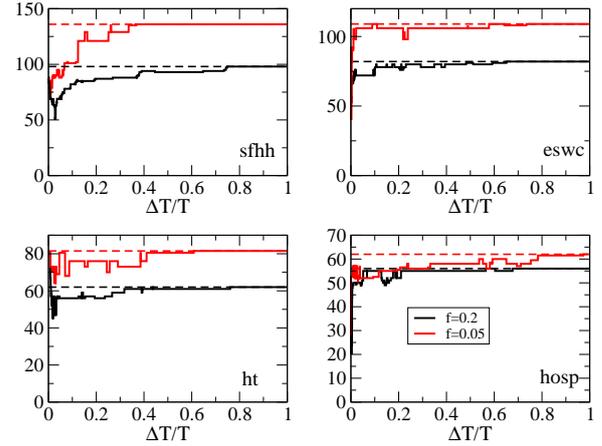}
  \end{center}
  \caption{Median degree {\em in the aggregated network} of the
    fraction $f=0.05$ (red) and the $f=0.2$ (black) nodes chosen by
    the $\textbf{K}$ strategy at $\Delta T$, vs $\Delta T/T$. The
    dashed horizontal lines mark the final values. }
  \label{fig:fhighestdegnode}
\end{figure}

In order to understand these results in more details, we have
considered the evolution with time of the nodes' properties. In
particular, we compare the largest degree nodes in the fully
aggregated network with the set of nodes ${\cal S}_{\textbf K}(\Delta
T)$, chosen by following the $\textbf{K}$ strategy on the training
window $[0,\Delta T]$. To this aim, we show in
Fig. \ref{fig:fhighestdegnode} the median of the degrees, {\em in the
  fully aggregated network}, of the nodes of ${\cal S}_{\textbf
  K}(\Delta T)$, as a function of $\Delta T$.  The median rapidly
reaches its final value, showing that, even for short $\Delta T$, the
set of immunized nodes ${\cal S}_{\textbf K}(\Delta T)$ has similar
properties (here the degree) than what would be obtained by taking
into account the whole dataset of length $T$.

\begin{figure}[t]
  \begin{center}
    \includegraphics[clip=true,trim=0cm 0.5cm 0cm 2.5cm,width=9cm]{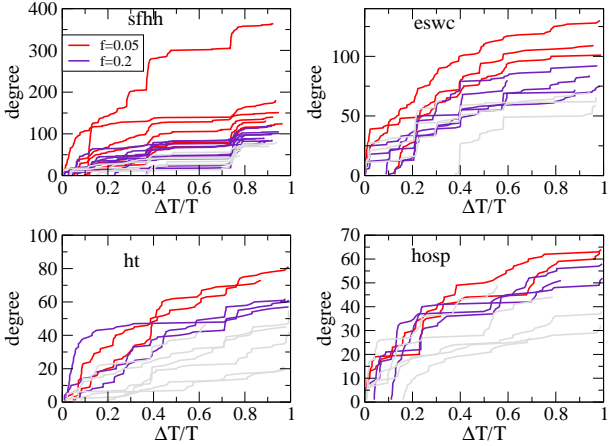}
  \end{center}
  \caption{Degree of nodes on the network aggregated over $\Delta T$,
    vs. $\Delta T/T$. The degrees of the fraction $f=0.05$ nodes with
    highest degree on the fully aggregated network are shown in red
    (only every third node is shown for clarity). The degree of the
    nodes ranked between $0.05N$ and $0.2N$ in the fully aggregated
    network are shown in blue (only every sixth node is shown). The
    evolution of the aggregated degree of a small number of other
    nodes is shown in grey for comparison. }
  \label{fig:degotime}
\end{figure}

Figure \ref{fig:degotime} moreover displays the evolution of the
degree aggregated over the training window $[0,\Delta T]$, as a
function of $\Delta T/T$, for several nodes.  The fraction $f=0.05$ of
nodes with the largest degree in the fully aggregated network are
ranked among the most connected nodes already for small training
windows $\Delta T$.  For $f=0.2$, the ranking fluctuates more and
takes longer to stabilize.  Overall, while the precise ordering scheme
of the nodes according to their degree is not entirely stable with respect to
increasing values of $\Delta T$, a coarse ordering is rather rapidly
reached: the nodes that reach a large degree at the end of the dataset
are rapidly ranked among the highest degree nodes, and the nodes that
in the end have a low degree are as well rapidly categorized as
such. This confirms the result of Fig. \ref{fig:fhighestdegnode} and
explains why the $\textbf{K}$ strategy reaches its best efficiency
even at short training windows for small $f$, and with a more gradual
saturation for larger fractions of immunized nodes.

The fact that high degree nodes are identified early on in the
information collection process comes here as a surprise: for a
temporal network with Poissonian events, all the information on the
relative importance of links and nodes is present in the data as soon
as the observation time is larger than the typical timescale of the
dynamics; this is however a priori not the case for the bursty
dynamics observed in real-world temporal networks. Various factors can
explain the observed stability in the ranking of nodes. On the one
hand, some nodes can possess some intrinsic properties giving them an
important a priori position in the network (for instance, nurses in a
hospital, or senior scientists in a conference) that ensure them a
larger degree than other nodes even at short times.  On the other
hand, the stability of the ranking could in fact be only temporary,
and due to the fact that nodes arriving earlier in the dataset have a
larger probability to gather a large number of contacts. In this case,
the observed stability of the ordering scheme could decrease for data
collected on longer timescales.

\begin{figure}[t]
  \begin{center}
    \includegraphics[clip=true,trim=0cm 0.5cm 0cm 2.5cm,width=9cm]{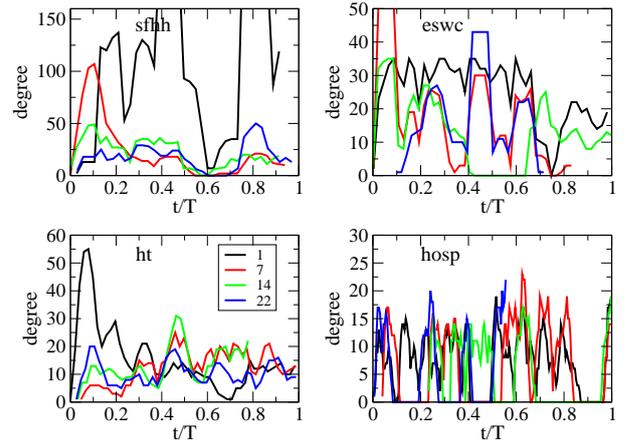}
  \end{center}
  \caption{Degree of the several nodes on networks aggregated over
    temporal windows of $400$ time steps $[t,t+400]$, vs $t/T$. The
    curves are colored according to the node's ranking in the network
    aggregated over $[0,T]$. }
  \label{fig:degotimeslice}
\end{figure}

Another reason for the saturation of the efficiency of the various
strategies is shown in Fig. \ref{fig:degotimeslice}, which displays
the evolution of the degree of some nodes in the network aggregated on
a temporal window $[t,t+400]$ versus $t/T$. This figure clearly shows
that the importance of different nodes varies strongly depending on
the initial time chosen for aggregation, and even a node with a large
degree in the network aggregated on $[0,\Delta T]$ can have
temporarily a small degree in a subsequent time window. The strong
variation in the degree of nodes at different times clearly limits the
efficiency not only of immunization strategies based on information
that is local in time, but even of strategies based on aggregated
information.

The main conclusion of our study is therefore twofold. On the one
hand, a limited amount of information on the contact patterns is
sufficient to design relevant immunization strategies; on the other
hand, the strong variation in the contact patterns significantly
limits the efficiency of any strategy based on importance ranking of
nodes, even if such deterministic strategies still perform much better
than the ``recent'' or ``weight'' protocols that are generalizations
of the ``acquaintance'' strategy. Moreover, strategies based on simple
quantities such as the aggregated degree perform as well as, or
better, than strategies based on more involved measures such as the
infection delay ratio defined in Sec.~\ref{sec:epid-models-numer}. We
also note that, contrarily to the case investigated by
\cite{Lee:2010fk}, the ``recent'' and ``weight'' strategies, which try
to exploit the temporal structure of the data, do not perform clearly
better than the simpler ``acquaintance'' strategy.  Such apparent
discrepancy might have various causes. In particular,
\cite{Lee:2010fk} consider spreading processes starting exactly at
$t=\Delta T$ while we average over different possible starting times.
The datasets used are moreover of different nature (\cite{Lee:2010fk}
indeed obtain contrasted results for different datasets) and have
different temporal resolutions. A more detailed comparison of the
different datasets' properties would be needed in order to fully
understand this point, as discussed for instance by \cite{Masuda13}.

\section{Acknowledgments}

RPS acknowledges ﬁnancial support from the Spanish MICINN, under
Project No. FIS2010-21781-C02-01, the Junta de Andaluc\'{\i}a, under
Project No. P09-FQM4682, and additional support through ICREA
Academia, funded by the Generalitat de Catalunya.
AB, CC and RPS are partly supported by FET project
MULTIPLEX 317532.


\end{document}